\documentclass[twocolumn,showpacs,preprintnumbers,amsmath,amssymb]{revtex4}

\usepackage{graphicx}
\usepackage{dcolumn}
\usepackage{bm}

\begin{document}

\title{Coherent exciton transport and trapping on long-range interacting cycles}
\author{Xin-Ping Xu$^{1,2}$}
\email{xuxp@mail.ihep.ac.cn}
\affiliation{%
$^1$Institute of Particle Physics, HuaZhong Normal University, Wuhan
430079, China \\
$^2$Institute of High Energy Physics, Chinese Academy of Science,
Beijing 100049, China
}%

\date{\today}

\begin{abstract}
 We consider coherent exciton transport modeled by continuous-time
quantum walks (CTQWs) on long-range interacting cycles (LRICs),
which are constructed by connecting all the two nodes of distance
$m$ in the cycle graph. LRIC has a symmetric structure and can be
regarded as the extensions of the cycle graph (nearest-neighboring
lattice). For small values of $m$, the classical and quantum return
probabilities show power law behavior $p(t)\sim t^{-0.5}$ and
$\pi(t)\sim t^{-1}$, respectively. However, for large values of $m$,
the classical and quantum efficiency scales as $p(t)\sim t^{-1}$ and
$\pi(t)\sim t^{-2}$. We give a theoretical explanation of this
transition using the method of stationary phase approximation (SPA).
In the long time limit, depending on the network size $N$ and
parameter $m$, the limiting probability distributions of quantum
transport show various patterns. When the network size $N$ is an
even number, we find an asymmetric transition probability of quantum
transport between the initial node and its opposite node. This
asymmetry depends on the precise values of $N$ and $m$. Finally, we
study the transport processes in the presence of traps and find that
the survival probability decays faster on networks of large $m$.
\end{abstract}
\maketitle
\section{Introduction}
In the past few years there has been a growing interest in
continuous-time random walks (CTRWs)~\cite{rn1,rn2,rn3}. The
particular surge of increasing interest can be partly attributed to
its close connection with the classical diffusion modeled by the
tight-binding model in condensed matter~\cite{rn4}. The quantum
mechanical analog of the classical diffusion process defined on
complex networks has also been studied with respect to the
localization delocalization transition in the presence of site
disorder~\cite{rn5}. In the literature, there are two main types of
quantum walks: continuous-time and discrete-time quantum
walks~\cite{rn6}. Discrete-time quantum walks evolve by the
application of a unitary evolution operator at discrete time
intervals, and continuous-time walks evolve under a time-independent
Hamiltonian~\cite{rn2}. It has been shown that on some graphs,
propagation between two properly chosen nodes is exponentially
faster in the quantum case~\cite{rn7}. In this respect, quantum
walks provide a good framework for the design of quantum algorithms
in the application of quantum computation~\cite{rn8}.

Here, we focus on continuous-time quantum walks (CTQWs). Previous
work have studied CTQWs on some particular graphs, such as, the
line~\cite{rn9,rad1}, cycle~\cite{rn10}, hypercube~\cite{rn11},
Cayley tree~\cite{rn12,rad2}, dendrimers~\cite{rn13} and other
regular networks with simple topology~\cite{rad3,rad4}. In
Ref.~\cite{rn14}, the authors studied the coherent exciton dynamics
on discrete rings under long-range step lengths distributed
according to $R^{-\gamma}$ ($\gamma\geqslant 2$). The strength of
the long-range interaction is a power law decay of the distance of
the nearest-neighboring lattice. They find that the long-range
interactions give no influence to the efficiency of the coherent
exciton transport~\cite{rn14}.

In this paper, we study the effect of long-range interactions on a
new network model, namely long-range interacting cycles (LRICs).
LRICs are constructed by connecting all the two nodes of distance
$m$ in the cycle graph (nearest-neighboring lattice). Therefore, the
network model has a symmetric structure and can be regarded as the
extensions of the cycle graph (nearest-neighboring lattice). The
newly added edges with large $m$ are long-range interactions and
serves as shortcuts in the nearest-neighboring cycle graph. A
detailed description of the network structure will be given in the
next section.

Since the structure of LRICs is completely symmetrical as the
nearest-neighboring cycle graph, we are able to analytically predict
the dynamical behavior of the coherent and incoherent transport. The
paper is organized as follows: In
Sec.~\uppercase\expandafter{\romannumeral 2} we give a description
to the structure of LRICs. In
Sec.~\uppercase\expandafter{\romannumeral 3}, we briefly review the
properties of CTQWs on general graphs. In
Sec.~\uppercase\expandafter{\romannumeral 4}, we derive analytical
results for LRICs and study the efficiency of the classical and
quantum transport by considering the scaling of the return
probability. Long time averages of the transition probabilities are
also studied in this section. In
Sec.~\uppercase\expandafter{\romannumeral 5}, we study trapping
process on LRICs. Conclusions and discussions are given in the last
part, Sec. \uppercase\expandafter{\romannumeral 6}.
\section{Topology and structure of LRICs}
\begin{figure}
\scalebox{0.8}[0.8]{\includegraphics{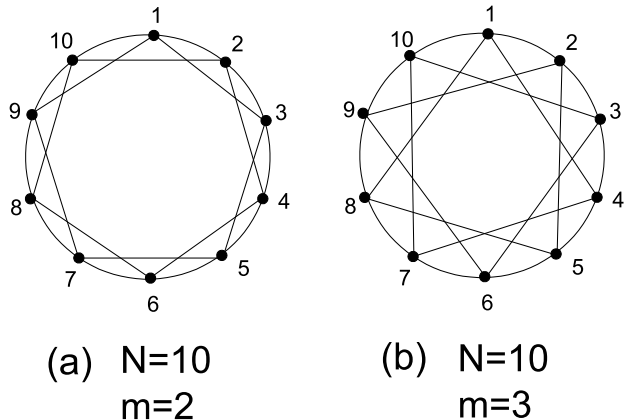}} \caption{(Color
online) Long-range interacting cycles $G(10,2)$ (a) and $G(10,3)$
(b). \label{fg1}}
\end{figure}

Long-range interacting cycles (LRICs) can be constructed as follows:
First, we construct a cycle graph of $N$ nodes where each node
connected to its two nearest neighbor nodes. Second, two nodes of
distance $m$ in the cycle graph are connected by additional bonds.
We continue the second step until all the two nodes of distance $m$
have been connected. Hence, the LRICs, denoted by $G(N,m)$, are
characterized by the network size $N$ and long-range interaction
parameter $m$. LRIC is a one-dimensional lattice with periodic
boundary conditions and all nodes of the networks have four bonds.
The structure of $G(10,2)$ and $G(10,3)$ is illustrated in
Fig.~\ref{fg1}.

It is interesting to note that all the LRICs have the same value of
connectivity $k=4$ and the parameter $m$ adjusts the interaction
range of the cycles, thus LRICs provide a good facility to study the
effects of long-range interaction on the transport dynamics.

\section{Coherent exciton transport on general graphs}
The coherent exciton transport on a connected network is modeled by
the continuous-time quantum walks (CTQWs), which is obtained by
replacing the Hamiltonian of the system by the classical transfer
matrix, i.e., $H=-T$~\cite{rn12,rn15}. The transfer matrix $T$
relates to the Laplace matrix by $T=-\gamma A$, where for simplicity
we assume the transmission rates $\gamma$ of all bonds to be equal
and set $\gamma \equiv 1$ in the following~\cite{rn12,rn15,rad5}.
The Laplace matrix $A$ has nondiagonal elements $A_{ij}$ equal to
$-1$ if nodes $i$ and $j$ are connected and $0$ otherwise. The
diagonal elements $A_{ii}$ equal to degree of node $i$, i.e.,
$A_{ii}=k_i$. The states $|j\rangle$ endowed with the node $j$ of
the network form a complete, ortho-normalised basis set, which span
the whole accessible Hilbert space. The time evolution of a state
$|j\rangle$ starting at time $t_0$ is given by $|j,t\rangle =
U(t,t_0)|j\rangle$, where $U(t,t_0)=exp[-iH(t-t_0)]$ is the quantum
mechanical time evolution operator. The transition amplitude
$\alpha_{k,j}(t)$ from state $|j\rangle$ at time $0$ to state
$|k\rangle$ at time $t$ reads $\alpha_{k,j}(t)=\langle
k|U(t,0)|j\rangle$ and obeys Schr\"{o}dinger¡¯s
equation~\cite{rn13,rn15,rad5}. Then the classical and quantum
transition probabilities to go from the state $|j\rangle$ at time
$0$ to the state $|k\rangle$ at time $t$ are given by
$p_{k,j}(t)=\langle k|e^{-tA}|j\rangle$ and
$\pi_{k,j}(t)=|\alpha_{k,j}(t)|^2= |\langle
k|e^{-itH}|j\rangle|^2$~\cite{rn12,rn15}, respectively. Using $E_n$
and $|q_n\rangle$ to represent the $n$th eigenvalue and
orthonormalized eigenvector of $H$, the classical and quantum
transition probabilities between two nodes can be written
as~\cite{rn12,rn13,rn15,rad5}
\begin{equation}\label{eq1}
p_{k,j}(t)=\sum_n e^{-tE_n}\langle k|q_n\rangle \langle
q_n|j\rangle,
\end{equation}
\begin{equation}\label{eq2}
\begin{array}{ll}
\pi_{k,j}(t)&=|\alpha_{k,j}(t)|^2=|\sum_n e^{-itE_n}\langle k|q_n\rangle \langle q_n|j\rangle|^2\\
&=\sum_{n,l} e^{-it(E_n-E_l)}\langle k|q_n\rangle \langle
q_n|j\rangle \langle j|q_l\rangle \langle q_l|k\rangle.
\end{array}
\end{equation}

For finite networks, $\pi_{k,j}(t)$ do not decay ad infinitum but at
some time fluctuates about a constant value. This value is
determined by the long time average of
$\pi_{k,j}(t)$~\cite{rn15,rad5}
\begin{equation}\label{eq3}
\begin{array}{ll}
\chi_{k,j}&=\lim_{T\rightarrow \infty}\frac{1}{T}\int_0^T
\pi_{k,j}(t)dt\\
&=\sum_{n,l}\langle k|q_n\rangle \langle q_n|j\rangle \langle j|q_l\rangle \langle q_l|k\rangle \\
&\  \ \  \ \times\lim_{T\rightarrow \infty}\frac{1}{T}\int_0^T e^{-it(E_n-E_l)}dt\\
&=\sum_{n,l}\delta_{E_n,E_l}\langle k|q_n\rangle \langle
q_n|j\rangle \langle j|q_l\rangle \langle q_l|k\rangle.
\end{array}
\end{equation}
where $\delta_{E_n,E_l}$ takes value 1 if $E_n$ equals to $E_l$ and
0 otherwise. Generally, to calculate $p_{k,j}(t)$, $\pi_{k,j}(t)$
and $\chi_{k,j}$ all the eigenvalues $E_n$ and eigenvectors
$|q_n\rangle$ are required. For some regular graphs, the eigenvalues
and eigenvectors can be analytically obtained. In the following
section, we find analytical results of the eigenvalues and
eigenstates for LRICs, and calculate these quantities according to
the above Equations.
\section{coherent transport on LRICs }
\subsection{Analytical results}
In the subsequent calculation, we restrict our attention on the
graph of long-range interacting cycles (LRICs). The network
organizes in a very regular manner and has a periodic boundary
condition. The Hamiltonian matrix $H$ of $G(N,m)$ ($m\geqslant 2$)
takes the following form,
\begin{equation}\label{eq4}
 H_{ij}=\langle i|H|j\rangle=\left\{
\begin{array}{ll}
4,   & {\rm if} \ i=j,\\
-1,   & {\rm if} \ i=j\pm 1,  \\
-1,   & {\rm if} \ i=j\pm m,  \\
 0,    & Otherwise.
\end{array}
\right.
\end{equation}
 And the Hamiltonian acting on the state $|j\rangle$ can be written as
\begin{equation}\label{eq5}
H|j\rangle=4|j\rangle -|j-1\rangle -|j+1\rangle -|j-m\rangle
-|j+m\rangle .
\end{equation}
The above Equation is the discrete version of the Hamiltonian for a
free particle moving on the cycles. Using the Bloch function
approach for the periodic system in solid state physics~\cite{rn16},
the time independent Schr\"{o}dinger¡¯s Equation reads
\begin{equation}\label{eq6}
H|\psi_n\rangle=E_n|\psi_n\rangle.
\end{equation}
The Bloch states $|\psi_n\rangle$ can be expanded as a linear
combination of the states $|j\rangle$ localized at node $j$,
\begin{equation}\label{eq7}
|\psi_n\rangle=\frac{1}{\sqrt{N}}\sum_{j=1}^N e^{-i\theta_n
j}|j\rangle.
\end{equation}
Substituting Eqs.~(\ref{eq5}) and (\ref{eq7}) into Eq.~(\ref{eq6}),
we obtain the eigenvalues (or energy) of the system,
\begin{equation}\label{eq8}
E_n=4-2\cos \theta_n -2 \cos(m\theta_n)
\end{equation}
The periodic boundary condition for the network requires that the
projection of the Bloch state on the state $|N+1\rangle$ equals to
that on the state $|1\rangle$, thus $\theta_n=2n\pi/N$ with $n$
integer and $n\in[1, N]$. Replacing $|q_n\rangle$ by the Bloch
states $|\psi_n\rangle$ in Eqs.~(\ref{eq1}), (\ref{eq2}) and
(\ref{eq3}), we can get the classical and quantum transition
probability
\begin{equation}\label{eq9}
p_{k,j}(t)=\frac{1}{N}\sum_n e^{-tE_n}e^{-i(k-j)2n\pi /N},
\end{equation}
\begin{equation}\label{eq10}
\begin{array}{ll}
\pi_{k,j}(t) &=|\alpha_{k,j}(t)|^2  \\
&=|\frac{1}{N}\sum_{n} e^{-itE_n}e^{-i(k-j)2n\pi /N}|^2,
\end{array}
\end{equation}
and the long time averages of $\pi_{k,j}(t)$ is given by
\begin{equation}\label{eq11}
\chi_{k,j}=\frac{1}{N^2}\sum_{n,l}\delta_{E_n,E_l}e^{-i(k-j)(n-l)2\pi/N}.
\end{equation}

Interestingly, when $k=j$, the transition probability is reduced to
the return probability, which means the probability of finding the
exciton at the initial node. In Ref.~\cite{rn17}, the authors use
the return probability to quantify the efficiency of the transport.
In the next subsection, we will analyze return probability and try
to compare the efficiency between the classical and quantum
transport. For our regular cycles, the return probability is
independent on the initial node. The average return probability can
be written as,
\begin{equation}\label{eq12}
p(t)=\frac{1}{N}\sum_j p_{j,j}(t)=\frac{1}{N}\sum_n e^{-tE_n},
\end{equation}
and
\begin{equation}\label{eq13}
\begin{array}{ll}
\pi(t)&=\frac{1}{N}\sum \pi_{j,j}(t)=|\alpha_{j,j}(t)|^2=|\bar{\alpha}(t)|^2  \\
&=|\frac{1}{N}\sum_{n} e^{-itE_n}|^2.
\end{array}
\end{equation}
Eqs.~(\ref{eq12}) and (\ref{eq13}) hold for finite networks. For
infinite networks, i.e., $N\rightarrow\infty$, the $\theta$ values
are quasi-continuous in Eq.~(\ref{eq8}). In the continuum limit, on
one hand, the eigenvalues of Eq.~(\ref{eq8}) can be rewritten as,
\begin{equation}\label{eq14}
E_m(\theta)=4-2\cos \theta -2\cos m\theta.
\end{equation}
on the other hand, the classical and quantum return probabilities in
Eqs.~(\ref{eq12}) and (\ref{eq13}) can be written as the following
integral form,
\begin{equation}\label{eq15}
p_m(t)=\frac{1}{2\pi}\int_{0}^{2\pi}exp(-tE_m(\theta))d\theta ,
\end{equation}
and
\begin{equation}\label{eq16}
\pi_m(t)=|\bar{\alpha}(t)|^2=|\frac{1}{2\pi}\int_{0}^{2\pi}exp(-itE_m(\theta))d\theta
|^2.
\end{equation}
\begin{figure}
\scalebox{1.0}[1.0]{\includegraphics{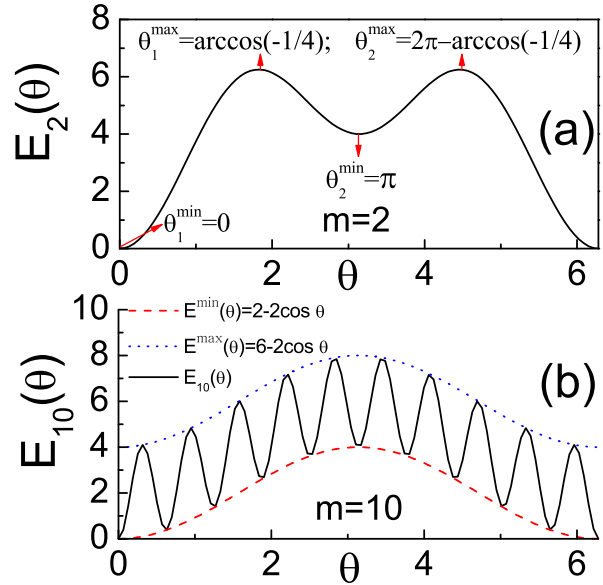}} \caption{(Color
online) Eigenvalues $E_m(\theta)$ versus $\theta$ for $m=2$ (a) and
$m=10$ (b). For $m=2$, there are two maxima and minima in
$[0,2\pi)$, which are indicated by the arrows in the plot (Fig.~2
(a)). For $m=10$, there are ten maxima and minima in $[0,2\pi)$, and
the maximal and minimal points are indicated as $E^{max}(\theta)$
and $E^{min}(\theta)$ (See dashed curves in Fig.~2 (b)).
\label{fg2}}
\end{figure}

Fig.~\ref{fg2} shows $E_m(\theta)$ versus $\theta$ for $m=2$ (a) and
$m=10$ (b). We note that $E_m(\theta)$ is an oscillatory function,
and there are more regular oscillations for large values of $m$. The
number of maxima (or minima) of $E_m(\theta)$ in the range
$[0,2\pi)$ is $m$. As we will show, these extreme points give
contributions to the integrals when we calculate the classical and
quantum efficiency in Eqs.~(\ref{eq15}) and (\ref{eq16}).
\subsection{Efficiency and scaling of the classical and quantum transport}
In this subsection, we consider the efficiency of the classical and
quantum transport. We calculate the integrals of Eqs.~(\ref{eq15})
and (\ref{eq16}) using the stationary phase approximation (SPA) (See
Appendix A). We find that the classical and quantum return
probabilities show different scaling behavior for small values and
large values of $m$.
\begin{figure}
\scalebox{1.0}[1.0]{\includegraphics{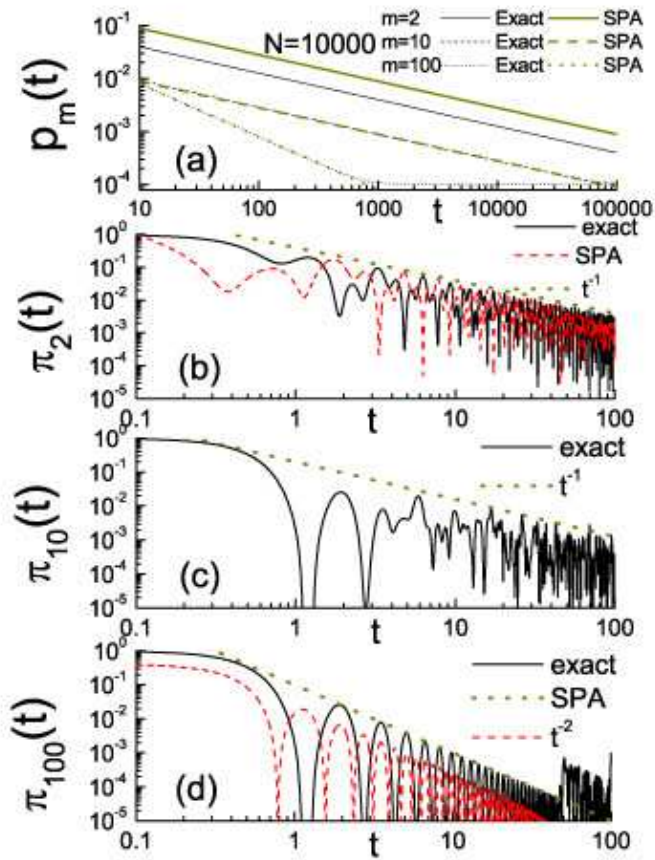}}
 \caption{(Color online)
Return probability $p_m(t)$ and $\pi_m(t)$ for different values of
$m$. (a)Classical return probabilities $p_m(t)$ for $m=2$, $m=10$
and $m=100$. The black curves are exact results obtained according
to Eq.~(\ref{eq12}), green curves are the analytical results
obtained using stationary phase approximation (SPA) (See
Eqs.~(\ref{b4}) and (\ref{b6}) in the Appendix). (b)Quantum return
probability $\pi_2(t)$. The solid curve is the exact result and the
dashed curve is the analytical result obtained using stationary
phase approximation (See Eq.~(\ref{c1}) in the Appendix).
(c)$\pi_{10}(t)$ versus $t$. The solid curve is the exact result and
the dotted line is the power law $t^{-1}$. (d)$\pi_{100}(t)$ versus
$t$. The solid curve is the exact result and the dashed curve is
analytical prediction according to Eq.~(\ref{eq19}). All the exact
results are obtained from LRICs of size $N=10000$. \label{fg3}}
\end{figure}

For small values of $m$, we get an asymptotical expression for the
classical $p_m(t)$,
\begin{equation}\label{eq17}
p_m(t)\approx \frac{1}{2m\sqrt{\pi t}}\sim t^{-0.5}.
\end{equation}
(See derivation in Appendix B). For large values of $m$, we also get
an approximate result (See Appendix B),
\begin{equation}\label{eq18}
p_m(t)\approx  \frac{1}{4\pi t}\sim t^{-1}.
\end{equation}
We note that $p_m(t)$ scales as $t^{-0.5}$ for small values of $m$,
however, for large $m$, the scaling becomes as $p_m(t)\sim t^{-1}$.

Quantum mechanically, we also find different scaling behavior of
$\pi_m(t)$ for small value and large value of $m$. For small values
of $m$, $\pi_m(t)$ is an oscillatory function multiplied by $1/t$
(See Eq.~(\ref{c1}) in Appendix C for $m=2$). For large values of
$m$, there is an approximate result given by Eq.~(\ref{c8}) in
Appendix C,
\begin{equation}\label{eq19}
\pi_m(t)\approx \frac{\sin^24t}{4\pi^2 t^2}\sim t^{-2}.
\end{equation}
Therefore, quantum transport of small $m$ displays the same scaling
behavior $\pi(t)\sim t^{-1}$ while transport of large $m$ shows
scaling $\pi(t)\sim t^{-2}$. It is interesting to note that, both
for the classical and quantum transport, the scaling exponent for
large $m$ is twice the exponent for small values of $m$. This is one
of the main conclusions in this paper.

In order to test the theoretical predictions, Fig.~\ref{fg3} shows
the classical and quantum return probabilities for LRICs of
$N=10000$ with $m=2$, $m=10$ and $100$. Fig.~\ref{fg3} (a) shows the
classical return probability. We note that $p_2(t)$ and $p_{10}(t)$
displays the same scaling $t^{-0.5}$, but for $m=100$ the scaling
becomes $p_{100}(t)\sim t^{-1}$. The results are in good agreement
with the analytical predictions of Eqs.~(\ref{eq17}) and
(\ref{eq18}). Fig.~\ref{fg3} (b) shows the quantum $\pi_m(t)$ for
$m=2$ and the analytical prediction of Eq.~(\ref{c1}) in Appendix C.
Both the results exhibit power law $\pi_2(t)\sim t^{-1}$. The same
scaling behavior ($t^{-1}$) is also observed for $m=10$ (See
Fig.~\ref{fg3} (c)). In Fig.~\ref{fg3} (d), we show $\pi_{100}(t)$
and the analytical result predicted by Eq.~(\ref{eq19}). Both the
results display the same scaling $t^{-2}$.
\subsection{Long time averages on finite networks}
\begin{figure}
\scalebox{1}[1]{\includegraphics{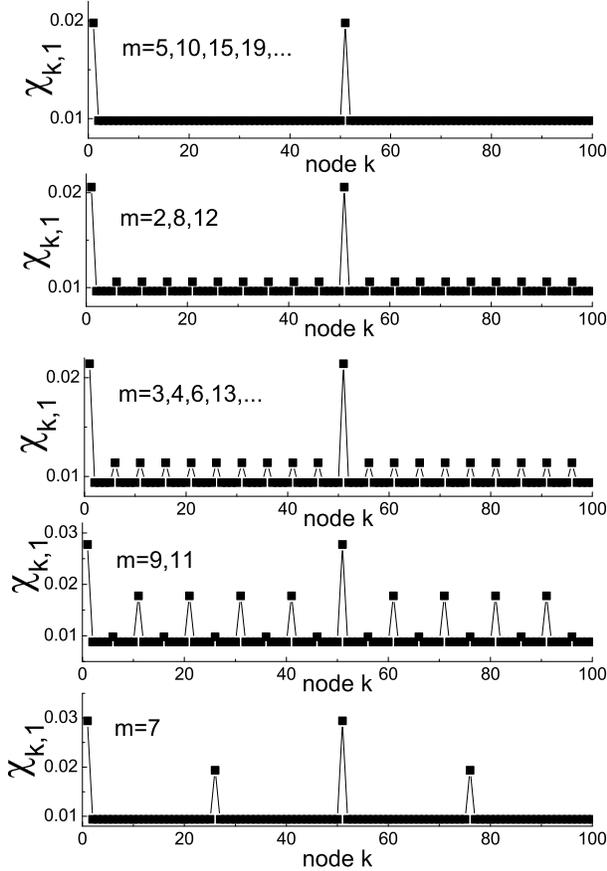}} \caption{ Long-time
averaged probability distribution $\chi{k,1}$ for CTQWs on networks
of size $N=100$ with different values of $m$. \label{fg4}}
\end{figure}
\begin{figure}
\scalebox{1}[1]{\includegraphics{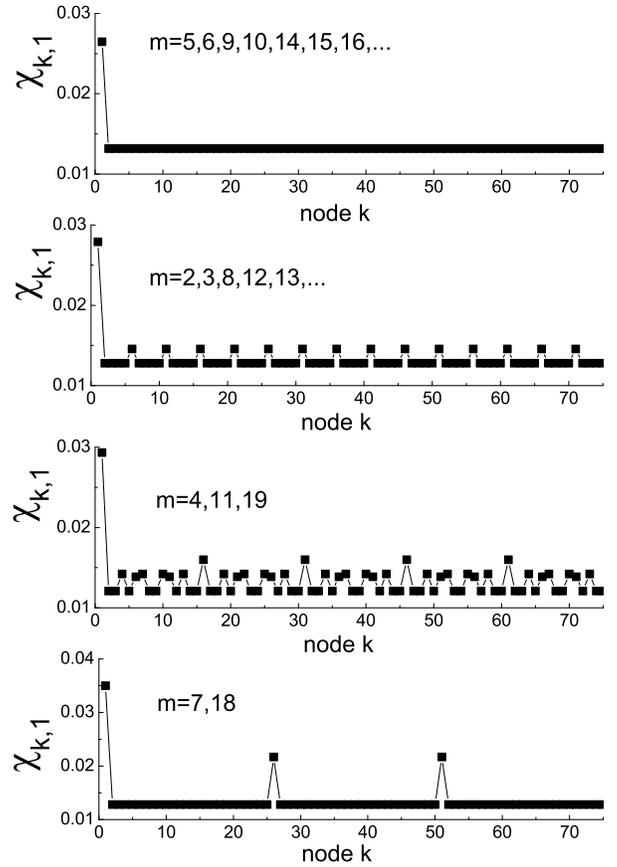}} \caption{Quantum
mechanical limiting probabilities $\chi{k,1}$ on networks of size
$N=75$ with different values of $m$. \label{fg5} }
\end{figure}
In this section, we consider the long time averaged transition
probabilities on finite networks. Classically, the long time liming
probabilities equal to the equip-partitioned probability
$1/N$~\cite{rn18}. Quantum mechanically, the long time averages of
the transition probabilities does not lead to equip-partition. For
LRICs, the long-time averaged probability is determined by
Eq.~(\ref{eq11}) but the distribution patterns are complex for
different network parameters $N$ and $m$. For the cycle graph
(nearest neighboring lattice), the limiting probability distribution
depends on the parity of the network size $N$. Fig.~\ref{fg4} shows
the distribution patterns of the limiting transition probability on
 networks of $N=100$ with various values of $m$. The initial excitation
is located at node $1$. As we can see, there are high probabilities
to find the exciton at the initial node $1$ and the opposite node
$51$, this feature is a natural consequence of the periodic boundary
condition of the graphs~\cite{rad5}. For odd-numbered networks $N\in
Odds$, there is a higher probability to find the initial node than
that at other nodes~\cite{rad5}. Fig.~\ref{fg5} shows the
distribution patterns for networks of $N=75$ with various $m$. The
patterns depend on the specific network parameters and there are
high probabilities to find the exciton at some particular nodes. We
also note that the patterns of $\chi_{k,1}$ are the same for some
different values of $m$, this feature can be explained by the
identical degeneracy distribution of the eigenvalues for different
values of $m$~\cite{rad5}.
\begin{figure}
\scalebox{0.8}[0.8]{\includegraphics{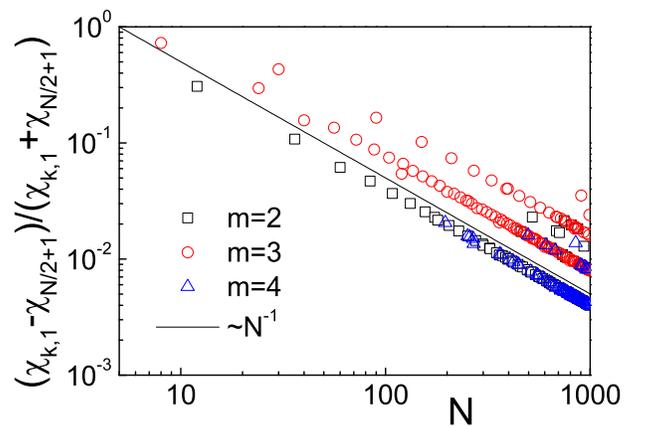}}
 \caption{(Color
online) $\Delta (1,N/2)\equiv
(\chi_{1,1}-\chi_{N/2+1,1})/(\chi_{1,1}+\chi_{N/2+1,1})$ as a
function of the network size $N$ for different values of $m$. The
solid line indicates the power law decay $\Delta(1,N/2)\sim N^{-1}$.
 \label{fg6}}
\end{figure}

It is worth mentioning that for even-numbered networks, there are
high probabilities to find the exciton at the initial node and
opposite node. For networks of $N=100$, we find the two
probabilities are exactly equal to each other for all the values of
$m$, i.e., $\chi_{1,1}=\chi_{51,1}$. However, for some other
even-numbered network size $N$, this is not true~\cite{rad5}. For
some particular values of $N$ and $m$, for instance $N=108$ and
$m=2$, the probability of finding the exciton at the initial node
differs from the probability of the finding the exciton at the
opposite node. Such asymmetry is small and not easy to be observed
from the limiting probability distributions~\cite{rad5}. To detect
such asymmetry of the probabilities, we define the quantity $\Delta
(1,N/2)\equiv
(\chi_{1,1}-\chi_{N/2+1,1})/(\chi_{1,1}+\chi_{N/2+1,1})$ as a
function of the network size $N$ for different values of
$m$~\cite{rad5}. The asymmetry is indicated by the nonzero of this
quantity while $\Delta (1,N/2)=0$ corresponds to identical values of
$\chi_{1,1}$ and $\chi_{N/2+1,1})$. A plot of $\Delta (1,N/2)$
versus $N$ for $m=2$, $m=3$ and $m=4$ are shown in Fig.~\ref{fg6}.
We find that the points break into several clusters, whereas some
clusters $\Delta (1,N/2)$ decreases with the network size $N$ as a
power law: $\Delta (1,N/2)\sim N^{-1}$~\cite{rad5}.

 As we have shown in Fig.~\ref{fg6}, the asymmetry appears at some
particular values of $N$ and $m$. However, we are unable to predict
which particular parameters of $N$ and $m$ are related to such
asymmetry. This is an interesting issue and similar phenomena is
also found in Ref.~\cite{rad5}.
\section{trapping on LRICs}
An important process related to random walk is
trapping~\cite{rn19,rn20}. Trapping problems have been widely
studied in the frame of physical chemistry, as part of the general
reaction-diffusion scheme~\cite{rn21}. Previous work has been
devoted to the trapping problem on discrete-time random
walks~\cite{rn22,rn23}. However, even in its simplest form, trapping
was shown to yield a rich diversity of results, with varying
behavior over different geometries, dimension, and time
regimes~\cite{rn23}. The main physical quantity related to trapping
process is the survival probability, which denotes the probability
that a particle survives during the walk in a space with traps.

In this paper, we consider trapping using the approach based on time
dependent perturbation theory and adopt the methodology proposed in
Ref.~\cite{rn24}. In Ref.~\cite{rn24}, the authors consider a system
of $N$ nodes and among them $M$ are traps ($M<N$). The trapped nodes
are denoted them by $m$, so that $m\in {\cal M}$. The new
Hamiltonian of the system is $H=H_0+i\Gamma$, where $H_0$ is the
original Hamiltonian without traps and $i\Gamma$ is the trapping
operator. $\Gamma$ has $m$ purely imaginary diagonal elements
$\Gamma_{mm}$ at the trap nodes and assumed to be equal for all $m$
($\Gamma_{mm}\equiv \Gamma >0$). See Ref.~\cite{rn24} for details.
The new Hamiltonian is non-hermitian and has $N$ complex eigenvalues
and eigenstates \{$E_l$, $|\Psi_l\rangle$\} ($l=1,2,...,N$). Then
the quantum transition probability is
\begin{equation}\label{eq20}
\pi_{k,j}(t)=|\alpha_{k,j}(t)|^2=|\sum_l e^{-itE_l}\langle k
|\Psi_l\rangle \langle \tilde{\Psi}_l|j\rangle|^2,
\end{equation}
where $\langle \tilde{\Psi}_l|$ ($l=1,2,...,N$) is the conjugate
eigenstates of the new Hamiltonian. In order to calculate
$\pi_{k,j}(t)$, all the complex eigenvalues and eigenstates \{$E_l$,
$|\Psi_l\rangle$\} ($l=1,2,...,N$) are required. Here, we
numerically calculate $\pi_{k,j}(t)$ by diagonalizing the
Hamiltonian H using the standard software package Mathematica 5.0.

Equation~(\ref{eq20}) depends on the initially excited node $j$. The
average survival probability over all initial nodes $j$ and all
final nodes $k$, neither of them being a trap node, is given by,
\begin{equation}\label{eq21}
\Pi_{M}(t)=\frac{1}{N-M}\sum_{j\not{\in} {\cal M}}\sum_{k\not{\in}
{\cal M}}\pi_{k,j}(t).
\end{equation}

For continuous-time random walks (CTRWs), we induce trapping
analogously as the CTQWs, where the new transfer matrix is modified
by the trapping matrix as $T=T_0-\Gamma$. The mean survival
probability analogous to Eq.~(\ref{eq21}) is
$P_{M}(t)=\frac{1}{N-M}\sum_{j\not{\in} {\cal
M}}\sum_{k\not{\in}{\cal M}}p_{k,j}(t)$~\cite{rn24}.
\begin{figure}
\scalebox{0.9}[0.9]{\includegraphics{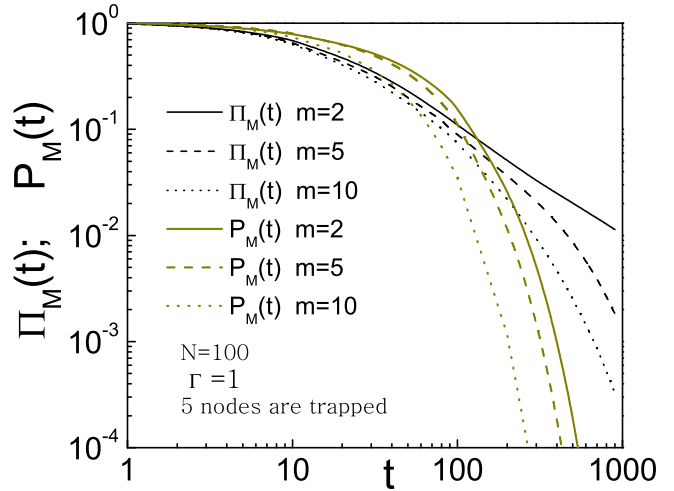}} \caption{(Color
online) Survival probabilities $\Pi_M(t)$ and $P_M(t)$ for LRICs of
$N=100$ and different values of $m$. In the calculation, five trap
nodes are randomly selected from the cycles and we set $\Gamma=1$.
The curves are averaged over distinct trapping realizations.
 \label{fg7}}
\end{figure}

Fig.~\ref{fg7} shows the quantum and classical survival
probabilities on LRICs of $N=100$ with $m=2$, $m=5$ and $m=10$. Five
trapped nodes are randomly selected from all the nodes and
$\Gamma=1$ is fixed in the numerical calculation. For each specific
trapping configuration, we calculate the survival probability and
average it over different configurations. As we can see from
Fig.~\ref{fg7}, both the quantum and classical survival
probabilities decays fast on LRICs with large values of $m$. This is
opposite to the case in Ref.~\cite{rn25} where long-range
interaction leads to a slower trapping of the excitation.
 \section{Conclusions and Discussions}
We have studied coherent exciton dynamics modeled by continuous-time
quantum walks (CTQWs) on long-range interacting cycles (LRICs). We
have shown that both the efficiency of the classical and quantum
transport display power laws, and the exponents for LRICs with large
values of $m$ are twice the exponents of LRICs with small values of
$m$. Theoretical calculation of the return probability using
stationary phase approximation supports this finding. In the long
time limit, the limiting probability distributions of quantum
transport show various patterns on finite size networks. When the
network size $N$ is an even number, we find an asymmetric transition
probability between the initial node and its opposite node. This
asymmetry depends on the precise values of $N$ and $m$. Finally, we
study trapping process on LRICs and find that long-range interaction
(large $m$) leads to a fast decay of the survival probability.

It is worth mentioning that the return probability displays
different scaling behavior for small values and large values of $m$.
However, we did not give a quantitative relation between the scaling
exponent and the parameter $m$. We only know the scaling behavior
for the thresholds of small and large $m$, the scaling behavior in
the medial region of $m$ is still unknown. In addition, the limiting
probability distributions show various patterns on finite size
networks. These patterns is a natural result of the interference
phenomena in coherent transport on finite systems. The asymmetry of
the limiting probability on even-numbered networks is also an
interesting and strange feature of quantum walks, which deserves our
further investigation~\cite{rn18,rn26}. The long-range interaction
in LRICs leads to a fast exciton trapping and is opposite to the
conclusions in Ref.~\cite{rn25}. We also note that the quantum
return probability in Ref.~\cite{rn14} scales the same behavior $\pi
(t)\sim t^{-1}$ for various long-range interactions ($\gamma
\geqslant 2$). The different behavior of trapping and transport
efficiency may be caused by the distinct type of long-range
interactions.
\begin{acknowledgments}
We thank Prof. Blumen (University of Freiburg) for useful
discussions. This work is supported by National Natural Science
Foundation of China under projects 10575042, 10775058 and MOE of
China under contract number IRT0624 (CCNU).
\end{acknowledgments}
\appendix
\section{The stationary phase approximation (SPA)}
Stationary phase approximation (SPA) is an approach for solving
integrals analytically by evaluating the integrands in regions where
they contribute the most~\cite{rn27,rn14,rad1}. This method is
specifically directed to evaluating oscillatory integrands, where
the phase function of the integrand is multiplied by a relatively
high value. Suppose we want to evaluate the behavior of function
$I(\lambda)$ for large $\lambda$,
\begin{equation}\label{a1}
I(\lambda)=\frac{1}{2\pi}\int e^{-\lambda f(x)}dx .
\end{equation}
The SPA asserts that the main contribution to this integral comes
from those points where $f(x)$ is stationary $[df(x)/dx\equiv
f'(x)\equiv 0]$. If there is only one point $x_0$ for which
$f'(x_0)=0$ and $d^2f(x)/dx^2|x_0\equiv f''(x_0)\neq 0$, the
integral is approximated asymptotically by,
\begin{equation}\label{a2}
I(\lambda)\approx \frac{1}{\sqrt{2\pi \lambda f''(x_0)}} e^{-\lambda
f(x_0)}.
\end{equation}
If there are more than one stationary points satisfy
$[df(x)/dx\equiv f'(x)\equiv 0]$, then the integral $I(\lambda)$ is
approximately given by the sum of the contributions [each being of
the form given in Eq.~(\ref{a2})] of all the stationary
points~\cite{rn14}.
\section{Calculation of the classical $p_m(t)$ using SPA}
We apply SPA to calculate the classical $p_m(t)$ in Eq.~
(\ref{eq15}). The stationary points of this integral satisfy
 $E'_m(\theta)=2\sin \theta +2m\sin m\theta =0$. The number of stationary
 points equals to $2m$ ($m$ maxima and $m$ minima, see
 Fig.~\ref{fg2}) in the range $\theta \in [0,2\pi)$. We denote $m$
 maxima stationary points as $\theta^{max}_i$ ($i=1,2,...,m$) and $m$
 minima stationary points as $\theta^{min}_i$ ($i=1,2,...,m$). Then
 the integral of Eq.~(\ref{eq15}) yields,
\begin{equation}\label{b1}
\begin{array}{ll}
p_m(t)&\approx \sum_{i=1}^m \frac{1}{\sqrt{2\pi t
E''_m(\theta^{min}_i)}} e^{-t E_m(\theta^{min}_i)} \\
& + \sum_{i=1}^m \frac{1}{\sqrt{2\pi t E''_m(\theta^{max}_i)}} e^{-t
E_m(\theta^{max}_i)},
\end{array}
\end{equation}
which is mainly determined by the small values of $E_m(\theta_i)$.
Considering $E_m(\theta^{min})\leqslant 4<E_m(\theta^{max})$,
contributions from the maximal stationary points in the above
equation is negligible. Therefore, $p_m(t)$ can be simplified as,
\begin{equation}\label{b2}
p_m(t)\approx \sum_{i=1}^m \frac{1}{\sqrt{2\pi t
E''_m(\theta^{min}_i)}} e^{-t E_m(\theta^{min}_i)} .
\end{equation}

For small values of $m$, the global minimum $E_m(\theta)$ at
$\theta=0$ is sufficiently separated from (smaller than) other local
minima. The sum in Eq.~(\ref{b2}) is mainly from the contribution at
the global minimum $\theta=0$, thus,
\begin{equation}\label{b3}
p_m(t)\approx \frac{1}{\sqrt{2\pi t E''_m(0)}} e^{-t E_m(0)} .
\end{equation}
Substituting the relation $E''_m(\theta)|_{\theta=0}=2+2m^2$ and
$E_m(0)=0$ into Eq.~(\ref{b3}), we get,
\begin{equation}\label{b4}
p_m(t)\approx \frac{1}{\sqrt{4\pi t(1+m^2)}}\approx
\frac{1}{2m\sqrt{\pi t}}\sim t^{-0.5}.
\end{equation}

For large values of $m$, the global minimum $E_m(\theta)$ at
$\theta=0$ is not sufficiently separated from (smaller than) other
local minima. The sum in Eq.~(\ref{b2}) contains contributions from
all the minimal stationary points. Noting that for large values of
$m$, the stationary points $\theta^{max}_i$ and $\theta^{min}_i$ are
approximately equidistant, i.e., $\theta^{min}_i\approx
2(i-1)\pi/m$, $\theta^{max}_i\approx (2i-1)\pi/m$ ($i=1,2,...,m$).
Therefore, we get the approximations: $E''_m(\theta^{min})\approx
2m^2$, $E''_m(\theta^{max})\approx -2m^2$. Thus Eq.~(\ref{b2}) can
be written as,
\begin{equation}\label{b5}
p_m(t)\approx  \frac{1}{2\sqrt{\pi t}} \Theta(t),
\end{equation}
where $\Theta(t)\equiv\frac{1}{m}\sum_{i=1}^m e^{-t
E_m(\theta^{min}_i)}$. In the continuum limit of large $m$,
$\Theta(t)$ equals to the integral $\frac{1}{2\pi}\int_0^{2\pi}
e^{-t E^{min}(\theta)}d\theta$, where $E^{min}(\theta)=2-2\cos
\theta$ (See the dashed curves in Fig.~\ref{fg2} (b)). We apply the
method of SPA again to evaluate this integral and find that the
contribution is mainly from the stationary point $\theta =0$, which
lead to $\Theta(t)\approx \frac{1}{2\sqrt{\pi t }}$. The classical
$p_m(t)$ of Eq.~(\ref{b5}) transforms into,
\begin{equation}\label{b6}
p_m(t)\approx  \frac{1}{4\pi t}\sim t^{-1}.
\end{equation}
\section{Calculation of the quantum $\pi_m(t)$ using SPA}
We calculate the integral of quantum return probability in
Eq.~(\ref{eq16}). We also find the integral displays different
scaling behavior for small values and large values of $m$.

For small values of $m$, we consider the case $m=2$, where there are
four stationary points: $\theta^{min}_1=0$, $\theta^{min}_2=\pi$,
$\theta^{max}_1=ArcCos(-1/4)$, $\theta^{max}_2=2\pi -ArcCos(-1/4)$
(See Fig.~\ref{fg2} (a)). The second-order derivations at these
points yield $E''(\theta^{min}_1)=10$, $E''(\theta^{min}_2)=6$ and
$E''(\theta^{max}_1)=E''(\theta^{max}_2)=-15/2$. The corresponding
spectral eigenvalues at the four points are $E(\theta^{min}_1)=0$,
$E(\theta^{min}_2)=4$ and
$E(\theta^{max}_1)=E(\theta^{max}_2)=25/4$. Using the method of SPA,
we obtain the integral of Eq.~(\ref{eq16}) for $m=2$ as,
\begin{equation}\label{c1}
\begin{array}{ll}
\pi_2(t)&=|\frac{1}{2\pi}\int_{0}^{2\pi}exp(-itE_m(\theta))d\theta
|^2\\
&\approx|\frac{1}{\sqrt{2\pi it\cdot10}}e^{-it\cdot0}+
\frac{1}{\sqrt{2\pi
it\cdot6}}e^{-it\cdot4}\\
&+ \frac{1}{\sqrt{2\pi it\cdot(-15)/2}}e^{-it\cdot25/4}|^2\\
&\approx \frac{1}{30\pi t}(12+\sqrt{15}\cos 4t-4\sqrt{5}\sin 9t/4\\
& \ \ -4\sqrt{3}\sin 25t/4)\sim t^{-1}.
\end{array}
\end{equation}
Therefore, the quantum mechanical efficiency scales as $\pi_2(t)\sim
t^{-1}$ for $m=2$. For other small values of $m$, the calculation is
analogous. The result is also an oscillatory function multiplied by
$1/t$. This suggests that the quantum transport of small $m$
displays the same scaling behavior $\pi(t)\sim t^{-1}$.

For the case of large $m$, the integral of Eq.~(\ref{eq16}) comes
from $2m$ stationary points,
\begin{equation}\label{c2}
\begin{array}{ll}
\pi_m(t)&\approx |\sum_{i=1}^m \frac{1}{\sqrt{2\pi it
E''_m(\theta^{min}_i)}} e^{-it E_m(\theta^{min}_i)} \\
&\ \ + \sum_{i=1}^m \frac{1}{\sqrt{2\pi it E''_m(\theta^{max}_i)}}
e^{-itE_m(\theta^{max}_i)}|^2 \\
&\approx |\frac{1}{2\sqrt{\pi it}}\cdot \frac{1}{m} \sum_{i=1}^m
e^{-it
E_m(\theta^{min}_i)}\\
&\ \ +\frac{1}{2\sqrt{-\pi it}}\cdot \frac{1}{m} \sum_{i=1}^m e^{-it
E_m(\theta^{max}_i)}|^2
\end{array}
\end{equation}
where in the last approximation $E''_m(\theta^{max})\approx -2m^2$
and $E''_m(\theta^{min})\approx 2m^2$ is applied. In the continuum
limit of $m\rightarrow \infty$, the sum in the above equation can be
written as the integral form,
\begin{equation}\label{c3}
\frac{1}{m} \sum_{i=1}^m e^{-it
E_m(\theta^{min}_i)}=\frac{1}{2\pi}\int_0^{2\pi}exp(-itE^{min}(\theta))d\theta,
\end{equation}
and
\begin{equation}\label{c4}
\frac{1}{m} \sum_{i=1}^m e^{-it
E_m(\theta^{max}_i)}=\frac{1}{2\pi}\int_0^{2\pi}exp(-itE^{max}(\theta))d\theta.
\end{equation}
Noting that $E^{min}(\theta)=2-2\cos \theta$ and
$E^{min}(\theta)=6-2\cos \theta$ (See the dashed curves in
Fig.~\ref{fg2} (b)), Eq.~(\ref{c2}) can be rewritten as,
\begin{equation}\label{c5}
\begin{array}{ll}
\pi_m(t)&\approx |\frac{1}{2\sqrt{\pi
it}}\frac{1}{2\pi}\int_0^{2\pi}exp(-it(2-2\cos \theta ))d\theta \\
& +\frac{1}{2\sqrt{-\pi
it}}\frac{1}{2\pi}\int_0^{2\pi}exp(-it(6-2\cos \theta ))d\theta |^2.
\end{array}
\end{equation}
For the two integrals in the above equation, we apply SPA again and
find that the contribution of this integral is mainly from two
stationary points $\theta =0$ and $\theta =\pi$. Thus
\begin{equation}\label{c6}
\frac{1}{2\pi}\int_0^{2\pi}exp(-it(2-2\cos \theta ))d\theta
\approx\frac{1}{2\sqrt{\pi it}}+\frac{1}{2\sqrt{-\pi it}}e^{-4it},
\end{equation}
and
\begin{equation}\label{c7}
\frac{1}{2\pi}\int_0^{2\pi}exp(-it(6-2\cos \theta ))d\theta
\approx\frac{1}{2\sqrt{\pi it}}e^{-4it}+\frac{1}{2\sqrt{-\pi
it}}e^{-8it}
\end{equation}
Substituting these relations into the Eq.~(\ref{c5}), we get,
\begin{equation}\label{c8}
\pi_m(t)\approx \frac{\sin^24t}{4\pi^2 t^2}\sim t^{-2}.
\end{equation}

\end{document}